\documentclass[useAMS,referee,usenatbib]{biom}

\usepackage[utf8]{inputenc}
\usepackage[pdftex]{graphicx}
\usepackage{epstopdf}
\usepackage{booktabs,siunitx}
\usepackage[toc]{appendix}
\usepackage{multirow}
\usepackage{tabularx}
\usepackage{array}
\usepackage{longtable}
\usepackage{textcomp} 
\usepackage{gensymb}
\usepackage[english]{babel}
\usepackage{url}
\usepackage{parskip}
\usepackage{dsfont}
\usepackage{amsmath}
\usepackage{amssymb}
\usepackage{enumitem}  
\usepackage{mathtools}
\usepackage{mathrsfs}
\usepackage[autostyle]{csquotes}
\usepackage{nicematrix}
\usepackage{caption}
\usepackage{natbib} 
\usepackage{calc}

\usepackage{setspace}
\usepackage{caption}
\usepackage{hyperref} 

\usepackage[normalem]{ulem}
\usepackage{xcolor}

\title[Bayesian Variable Selection on Small Sample Trial Data via Adaptive Posterior-Informed Shrinkage Prior]{Bayesian Variable Selection on Small Sample Trial Data via Adaptive Posterior-Informed Shrinkage Prior}

\author{Lingxuan Kong$^{1,*}$\email{lingxuko@umich.edu}, 
Yumin Zhang$^{2,**}$\email{Yumin\_Zhang@vrtx.com},
Chenkun Wang$^{2,***}$\email{Chenkun\_Wang@vrtx.com}, and 
Yaoyuan Vincent Tan$^{2,****}$\email{Vincent\_Tan@vrtx.com} \\
$^{1}$Department of Biostatistics, University of Michigan, Ann Arbor, Michigan, 48105, U.S.A. \\
$^{2}$Vertex Pharmaceutics Inc., Boston, 02210, Massachusetts, U.S.A.}

\begin{document}



\date{{\it Received Dec} 2022. {\it Revised 000} 0000.  {\it Accepted 000} 0000.}



\pagerange{\pageref{firstpage}--\pageref{lastpage}} 
\volume{0}
\pubyear{0000}
\artmonth{000}




\label{firstpage}


\begin{abstract}
Identifying variables associated with clinical endpoints is of much interest in clinical trials. With the rapid growth of cell and gene therapy (CGT) and therapeutics for ultra-rare diseases, there is an urgent need for statistical methods that can detect meaningful associations under severe sample-size constraints. Motivated by data-borrowing strategies for historical controls, we propose the Adaptive Posterior-Informed Shrinkage Prior (APSP), a Bayesian approach that adaptively borrows information from external sources to improve variable-selection efficiency while preserving robustness across plausible scenarios. APSP builds upon existing Bayesian data borrowing frameworks, incorporating data-driven adaptive information selection, structure of mixture shrinkage informative priors and decision making with empirical null to enhance variable selection performances under small sample size. Extensive simulations show that APSP attains better efficiency relative to traditional and popular data-borrowing and Bayesian variable-selection methods while maintaining robustness under linear relationships. We further applied APSP to identify variables associated with peak C-peptide at Day 75 from the Clinical Islet Transplantation (CIT) Consortium study CIT06 by borrowing information from the study CIT07.
\end{abstract}

%

\begin{keywords}
Bayesian hierarchical modeling, Bayesian variable selection, adaptive data borrowing, shrinkage priors, CIT06, CIT07
\end{keywords}


\maketitle


%

\section{Introduction}

In drug discovery, the main objective of the sponsor is to find the drug or therapy that is efficacious and safe for the indicated disease or condition. Once approved, the clinician or practitioner shifts their interest to answering the question ``How likely or how well will the therapy or drug work on my patient?'' The scientists also hope to gain more knowledge on the working mechanisms behind the drug or therapy. To address these concerns, the researchers usually set up real-world evidence studies with the aim of exploring and focusing on the variables that could potentially be associated with the clinical endpoint. 

Within the context of traditional small molecular drugs for relatively high incidence rate diseases, there will be sufficient patients recruited to allow efficient identification of the variable-outcome associations, where sample size is not a concern from the aspect of statistical power. Recently, the blooming growth in the development and use of cell and gene therapies (CGT) to treat rare diseases, for example, blood cancer, sickle cell disease, and Type 1 Diabetes (T1D), brings new challenges in the identification of the variable-outcome associations. The hallmark of CGT is the overwhelming efficacy and small sample size in both trial and target population. With small sample sizes, identifying variables associated with clinical endpoint faces a huge challenge in statistical power as there is too little information provided by the data itself to draw any useful statistical conclusions. When the total number of variables approaches or greater than the sample size, direct modeling approach is no longer feasible \citep{Freedman1983, Hastie2001}. To address this, various variable selection methods based on local models, such as Least Absolute Shrinkage and Selection Operator (LASSO), spike-and-slab priors, and horseshoe priors, have been developed and widely applied \citep{Hans2009, Ishwaran2005, Carvalho2010}. These methods perform variable selection by introducing penalties for including additional covariates or by employing shrinkage priors that encourage regression coefficients to shrink toward zero if the effect size is small. However, it is well documented that such methods may yield biased estimates and tend to be too conservative to miss covariates with small but meaningful effects, for example, the ``beta-min'' condition in LASSO \citep{vandeGeer2011, Hans2009}.

As an alternative to shrink estimates under limited information, data borrowing techniques could be used to incorporate historical datasets or similar clinical trials—to improve estimation efficiency, as we believe in the similarities among the control groups or the potential shared working mechanism of the same therapy on multiple populations. The Bayesian framework is particularly well suited for this purpose, as external data can be naturally incorporated through informative priors, while internal data serve to update our beliefs about the effects of covariates on the endpoint. Bayesian data borrowing methods have demonstrated advantages in enhancing statistical power and reducing required sample sizes. Notably, power priors, commensurate priors and meta-analytic predictive priors have been widely used for integrating historical data into current studies, and their statistical properties have been extensively studied \citep{Hobbs2011, Hobbs2012, Ibrahim2015, Schmidli2014}. More recently, adaptive data borrowing strategies—such as self-adaptive mixture priors and elastic priors—have been proposed to improve flexibility and performance \citep{Jiang2023, Yang2023}. These approaches have been particularly successful in the design of basket trials, where treatment effect information is shared across subgroups, and the degree of borrowing is determined by the estimated similarity between datasets \citep{Ouma2022, Zheng2022, Zheng2023}. A comprehensive overview of data borrowing methodologies in the context of basket trials is provided by \citet{Zhou2024}. 

Despite these advancements, existing data borrowing methods often impose strong assumptions regarding the similarity between external and internal data, or they uniformly borrow the same amount of information across all variables \citep{Ibrahim2015}. In many cases, commensurability is pre-specified by measuring similarity between external data and internal data, which lacks flexibility and stability under small sample sizes \citep{Hobbs2012}. Meta-analytic predictive priors are more robust to discrepancies between external data and internal data, but the local sample size limits its efficiency in variable selection \citep{Schmidli2014}. These limitations reduce the applicability of current approaches to variable selection in small-sample settings, where discrepancies between internal and external data may lead to misleading inference. In particular, over-reliance on external data can result in both false positives and false negatives.

In this work, we propose a novel Adaptive Posterior-Informed Shrinkage Prior (APSP) framework that takes the advantage of data borrowing and shrinkage properties of existing methods to enhance variable selection efficiency and robustness on small sample data. APSP adaptively borrows meaningful information from external data and flexibly deal with the possible discrepancies between external data and internal data. Simultaneously, it achieves shrinkage through the mixture structure of informative prior and shrinkage prior, which shares similarity in structure with classic spike-and-slab priors and robust meta-analytic predictive prior but more flexible and robust to small sample size. Conceptually, APSP can be viewed as a smart decision between fitting a local shrinkage model or updating prior beliefs using informative external data. Compared to existing approaches, the proposed method offers several key advantages: (i) variable-specific information borrowing, (ii) regularization through shrinkage estimation, and (iii) extensibility to multiple external data sources via a tailored data-source selection mechanism. The remainder of the paper is organized as follows. In Section 2, we present the proposed APSP framework in detail. Section 3 summarizes extensive simulation results that demonstrate its performance. In Section 4, we apply the method to identify variables associated with a clinical endpoint using data from the Clinical Islet Transplantation (CIT) protocol version 6 and 7, two Type 1 Diabetes clinical trials conducted by National Institutes of Health (NIH). Sections 5 concludes the study and discusses possible extensions and future directions.

\section{Method}
\label{s:model}

\subsection{Notations}
Suppose we have two datasets with different sample sizes, the dataset with smaller sample size is referred to as the internal data and the dataset with relatively larger sample size is referred as the external data. We let $Y$ denote the clinical endpoint and $\mathbf{X}$ denote the variable that may be associated with the clinical endpoint. Then we can further denote the internal data with sample size $n^I$ as $\mathcal{D}^{I}=\{Y_i,\mathbf{X}_i\}_{i=1}^{n^I}$ and external data with sample size $n^E$ as $\mathcal{D}^{E}=\{Y_i,\mathbf{X}_i\}_{i=1}^{n^E}$. Let $\mathbf{X}^{E}$ and $\mathbf{X}^{I}$ denote the variables collected in external data and internal data respectively with $K^{E}$ and $K^{I}$ representing the number of variables as well as the dimension of covariate space. We further let $\mathbf{X}^{E,1}$ and $\mathbf{X}^{I,1}$ denote the variables that are associated with the clinical endpoint in external data and internal data respectively, and $\mathbf{X}^{E,0}$ and $\mathbf{X}^{I,0}$ denote the spurious variables in external data and internal data correspondingly. Then we can write the associations between $Y$ and $\mathbf{X}^{1}$ as $Y^{I}\sim g(\mathbf{X}^{I,1}\mid \boldsymbol{\beta}^{I,1})$ and $Y^{E}\sim g(\mathbf{X}^{E,1}\mid \boldsymbol{\beta}^{E,1})$, where $g(\cdot)$ serves as the link function and $\boldsymbol{\beta}^{1}$ can be interpret as the average effect of $\mathbf{X}^{1}$ on $Y$. We assume $g(\cdot)$ is the identical link in the following sections as we aim at identifying covariates that are related to a continuous clinical endpoint in our application.

\subsection{Assumptions}
In the existing data borrowing literature, relatively strong assumptions are often imposed on the regression coefficients, for example, the assumption of full exchangeability between internal and external datasets. Acknowledging the potential discrepancies between the internal coefficients $\boldsymbol{\beta}^{I}$ and the external coefficients $\boldsymbol{\beta}^{E}$, we relax such strong assumptions and instead adopt a milder condition to ensure that data borrowing remains meaningful in the scenarios for which the proposed method is designed.

\textit{Assumption 1}: Non-empty shared signal covariate space $\mathcal{X}^{I,1}\cap \mathcal{X}^{E,1} \neq \emptyset$ with $\mathbf{X}^I \in \mathcal{X}^I=\mathcal{X}^{I,1} \cup \mathcal{X}^{I,0}$, $\mathbf{X}^{E} \in \mathcal{X}^{E}=\mathcal{X}^{E,1} \cup \mathcal{X}^{E,0}$.\label{assumption1}

\textit{Assumption 2}: Both $\mathbf{X}^{I}$ and $\mathbf{X}^{E}$ are from the same distribution family and can be scaled.

\textit{Assumption 3}: Additive effect of both $\mathbf{X}^{I,1}$ and $\mathbf{X}^{E,1}$ on $Y^{E}$ and $Y^{I}$ respectively, with possibly varying effect sizes. 

Assumption 1 implies that the internal and external datasets share some variables that are truly associated with the clinical endpoint. This overlap justifies borrowing information from the external data to enhance parameter estimation in the internal dataset. In our simulation study results, we demonstrate that the proposed method can still achieve robust variable selection by adaptively down-weighting the impact of information borrowed from the external data, even when Assumption 1 is violated. Assumption 2 further ensures the information borrow argument works as $\boldsymbol{\beta}$ is interpreted as the marginal effects of unit change of $\mathbf{X}$ or the between group differences. If the variables in $\mathbf{X}^{E}$ and $\mathbf{X}^{I}$ are coded in different ways, then the borrowing of $\boldsymbol{\beta}$ will not work or even introduce bias in estimation. 
Assumption 3 guarantees that simple models with linear combinations of $\mathbf{X}$ can capture the associations between variables and clinical endpoint, and $\boldsymbol{\beta}$ is valid for measuring the covariate's effect size. Here we define $\boldsymbol{\beta}$ as:

\begin{longequation}
    \begin{array}{l}
        \beta_{j} = \mathbb{E}(Y\mid X_j+\Delta_{X_j},\mathbf{X}_{-j})-\mathbb{E}(Y\mid X_j,\mathbf{X}_{-j})
    \end{array}
    \label{eq:beta}
\end{longequation}

Here $\Delta_{X_j}$ represents a general change of $X_j$, which can be unit change for continuous variables and group change for categorical variables. Under assumption 3, we adopt $Y=g(\sum_{j=1}^{K}\beta_jX_j)$ as working models for both internal and external data.

\subsection{Adaptive Posterior-Informed Shrinkage Prior (APSP)}
Similar to mainstream data borrowing approaches, our method aims to leverage useful information from external data to improve parameter estimation in internal datasets, particularly when small internal sample sizes lead to unstable estimates under both Bayesian and frequentist frameworks. APSP modeling contains three major steps: first, obtain information of $\boldsymbol{\beta}^{E}\mid \mathcal{D}^{E}$ and the working structure of APSP; secondly, construct Posterior-Informed Shrinkage Prior (APSP) based on the information from $\mathcal{D}^{E}$; finally, estimate $\boldsymbol{\beta}^{I}\mid \mathcal{D}^{I},\boldsymbol{\beta}^{E}$ through modeling the internal data. Under this modeling process, both the information of parameter $\boldsymbol{\beta}^{E}$ and the signal pattern $\mathcal{X}^{E}$ is borrowed from the external data. 

As the first step, multiple models may be considered to obtain $\boldsymbol{\beta}^{E}\mid \mathcal{D}^{E}$ and the working structure of APSP. Here we consider a class of shrinkage modeling approaches when fitting local models to the external data, based on the following rationale: First, the shrinkage estimates provide biased, yet meaningful parameter estimates by producing posterior means of $\boldsymbol{\beta}$. Second, the shrinkage methods help screen out the potential spurious variables in the external data. When the external data has a much larger sample size compared to the internal data, carrying over such nuisance estimates into local model may lead to higher false positive rates. Borrowing trusted signals confirmed by the external data is a more robust approach compared to use non-shrinkage methods. For example, the external model can be fitted using the framework described by \citep{Ishwaran2005}:
\begin{equation}
    \begin{split}
        Y_{i}^{E}\mid \boldsymbol{\beta}^{E}, \sigma^2 &\sim \mathcal{N}(\mathbf{X}^{T}_{i}\boldsymbol{\beta}^{E}, \sigma^2),\ i=1,\cdots,n^{E}\\
        \beta_{k}^{E} \mid \gamma_{k}^{E} &\sim \gamma_{k}^{E}\mathcal{N}(0, \tau_{k}^2) + (1-\gamma_{k}^{E})\mathcal{N}(0, m_{k}\tau_{k}^2),\ k=1,\cdots, K^{E} \\
        \gamma_{k}^{E}\mid \pi_{k}^{E} &\sim Bernoulli(\pi_{k}^{E}) \\
        \pi_{k}^{E} &\sim Beta(a_{\pi}^{E}, b_{\pi}^{E}) \\
        \sigma^2 &\sim Gamma(a_{\sigma}, b_{\sigma})
    \end{split}
    \label{eq:SSP}
\end{equation}

We assign non-informative priors to the variance term $\sigma^2$ and the latent selection probability $\pi_k^{E}$, where $a_{\cdot}^{E}$ and $b_{\cdot}^{E}$ are pre-specified hyperparameters. In the spike-and-slab formulation (\ref{eq:SSP}), the slab component is represented by a normal distribution with a small precision parameter $\tau_k^2$, and the spike component is modeled using a normal distribution with precision parameter $m_k\tau_k^2$, where the pre-defined parameter $m_k$ is set to a large value to ensure the distribution effectively degenerates at zero. Due to the mixture prior nature of spike-and-slab prior, the posterior mean of $\boldsymbol{\beta}^{E} \mid \mathcal{D}^{E}$ is a biased estimator of $\boldsymbol{\beta}^{E}$. With enough external data sample or strong signal, the posterior mean of $\pi_{k}^{E}$ approximates 1, where the posterior distribution of $\boldsymbol{\beta}^{E} \mid \mathcal{D}^{E}$ converges to the posterior distribution under non-informative prior. Another advantage of applying shrinkage methods is, we have more tools in assessing the relative importance of variables and decide the working structure of APSP based on the external data. A classical and widely used way to select important variable is to pick the variables with $95\%$ confidence interval or credible interval not across 0, while in the spike-and-slab prior model, we can evaluate the proportion of zero and non-zero $\beta$ sampled from the posterior distribution of spike and slab part respectively by looking at the posterior samples of $\gamma_{k}^{E}$, which is denoted as $\gamma_{k}^{E}\mid \mathcal{D}^{E}$. This proportion is estimated by $\mathbb{E}\gamma_{k}^{E}\mid \mathcal{D}^{E}$, known as the posterior inclusion probability (PIP) of $\beta_{k}^{E}$, and $X_{k}^{E}$ will be identified as signal if $\text{PIP}$ exceeds a pre-defined threshold. PIP helps us decide the working structure of APSP as we want to borrow the information of variables that have higher probability of having true association with the clinical endpoint, and avoid introduce biased estimate of variables that have high probability of being spurious variables into the estimated of $\boldsymbol{\beta}^{I}$. The combination of $\widehat{\boldsymbol{\beta}}^{E}$ and $\widehat{\text{PIP}}^{E}$ reveals our best guess on the effect size and signal pattern based on the external data.

With the information from external data, we construct the APSP as follows:
\begin{equation}
    \begin{split}
        \beta_{k}^{I} \mid \gamma_{k}^{I}, \tau_{\beta_{k}}^2 &\sim \gamma_{k}^{I}\bigg[\delta_k\underbrace{\mathcal{N}(\beta_{k}^{E}, \tau_{\beta_{k}}^2var(\beta_{k}^{E}))}_{Informative\ Prior}+(1-\delta_k)\underbrace{\mathcal{N}(0, \tau_{\beta_{k}}^2\tau_k^2)}_{Slab}\bigg]\\
        &
        +(1-\gamma_{k}^{I})\underbrace{\mathcal{N}(0, m_k\tau_{k}^2)}_{Spike},\ k=1,\cdots, K^{I}
    \end{split}
    \label{eq:APSP}
\end{equation}

APSP maintains a mixture structure of non-zero prior component and shrinkage prior component, similar to the spike-and-slab prior, and the information we obtained from the external data is incorporated by adding the modified posterior distribution of $\beta_{k}^{E}$ into the non-zero prior component, which is the $\mathcal{N}(\beta_{k}^{E}, \tau_{\beta_{k}}^2var(\beta_{k}^{E}))$ term. The classical non-informative slab component will remain in the prior if no information from external data can be incorporated, where the posterior-informed shrinkage prior will be reduced to a spike-and-slab prior. Pre-defined parameters of the spike component and the slab component are the same as the spike-and-slab prior formulation (\ref{eq:SSP}), where $\tau_k^2$ is a relatively small precision parameter and $m_k$ is used to let $\mathcal{N}(0, m_k\tau_{k}^2)$ serve as the spike component. $\delta_{k}\in \{0,1\}$ controls whether external information is borrowed, and $\delta_k$ can be estimated by $\widehat{\text{PIP}}_{k}^{E}>C^{E}_{k}$ or decided by internal data by assigning a hyperparameter to $\delta_{k}$, where $C^{E}_{k}$ is a pre-defined threshold of including the modified posterior distribution of $\beta_{k}^{E}$ into the APSP. Under the small sample scenario, the internal data may not be able to yield a stable posterior distribution of $\delta_k$, and we choose to estimate $\delta_k$ by $\widehat{\delta}_k=I(\mathbb{E}\gamma_{k}^{E}\mid \mathcal{D}^{E}>C^{E}_{k}), C^{E}_{k}=0.5$. In this case, $\boldsymbol{\gamma}^{E}\mid \mathcal{D}^{E}$ indicates whether posterior distribution of $\beta_{k}^{E}$ should be borrowed based on the external data, and $\boldsymbol{\beta}^{E}\mid \mathcal{D}^{E}$ informs us $\boldsymbol{\beta}^{I}$ if the external data supports incorporating external information into local model fitting process. If $\widehat{\delta}_k=0$, then based on external data, we believe that $\mathbf{X}_{k}^{E}$ has limited ability in explaining $Y$ compared to any variable $\mathbf{X}_{k'}^{E}$ with $\widehat{\delta}_{k'}=1$, which also means that the information we obtained from $\mathcal{D}^{E}$ is that $\mathbf{X}_{k}^{E}$ is less likely to be associated with $Y$, and hence, minimal to no borrowing occurs and applying a non-informative slab prior has the same efficiency and is more robust.

Under the Bayesian framework, if internal data and external data confirm and support the borrowing of $\boldsymbol{\beta}^{E}\mid \mathcal{D}^{E}$, then $\boldsymbol{\beta}^{I}\mid \mathcal{D}^{I},\boldsymbol{\beta}^{E}$ will have same posterior mean and smaller posterior variance compared to $\boldsymbol{\beta}^{I}\mid \mathcal{D}^{I}$. On the other hand, if internal data and external data do not support borrowing, this method is also capable of dealing with signals that only exist in internal data as the non-informative slab component will be updated to a posterior distribution with non-zero mean. The hyperparameter $\tau_{\beta_{k}}^{2}$ ensures flexible adjustment to the possible discrepancy between external data and internal data.

As the last step, with the constructed posterior-informed shrinkage prior, we fit a local hierarchical model based on the internal data:
\begin{equation}
    \begin{split}
        Y_{i}^{I}\mid \boldsymbol{\beta}^{I}, \sigma^2 &\sim \mathcal{N}(\mathbf{X}^{T}_{i}\boldsymbol{\beta}^{I}, \sigma^2),\ i=1,\cdots,n^{I}\\
        \beta_{k}^{I} \mid \gamma_{k}^{I}, \tau_{\beta_{k}}^2 &\sim \gamma_{k}^{I}\bigg[\widehat{\delta}_k\mathcal{N}(\widehat{\beta}_{k}^{E}, \tau_{\beta_{k}}^2\widehat{var}(\beta_{k}^{E}))+(1-\widehat{\delta}_k)\mathcal{N}(0, \tau_{\beta_{k}}^2\tau_{k}^2)\bigg]\\
        &+ (1-\gamma_{k}^{I})\mathcal{N}(0, m_k\tau_{k}^2),\ k=1,\cdots, K^{I}\\
        \gamma_{k}^{I} \mid \pi_{k}^{I} &\sim Bernoulli(\pi_{k}^{I})\\
        \pi_{k}^{I} &\sim Beta(a_{\pi}^{I}, b_{\pi}^{I}) \\
        \tau_{\beta_{k}}^2 & \sim Gamma(a_{\tau_{\beta_{k}}^2}^{I}, b_{\tau_{\beta_{k}}^2}^{I})\\
        \sigma^2 &\sim Gamma(a_{\sigma}, b_{\sigma})
    \end{split}
    \label{eq:APSPModel}
\end{equation}

Similar to the external model, we assign non-informative prior to $\sigma^2$ and $\tau_{\beta_{k}}^2$, with $a_{\cdot}^{I}$ and $b_{\cdot}^{I}$ being the pre-specified hyperparameters. We do not distinguish the hyperparameters assigned to $\sigma^2$ and parameters $\tau_{k}^2, m_k$ in external model and internal model as under the given model specification, there is no impact if we use shared hyperparameters, and $m_k$ and $\tau_{k}^2$ are pre-defined standard parameters in spike-and-slab priors, invariant to data source and $k$. APSP can be implemented conveniently using existing MCMC sampling algorithms such as Gibbs sampling or Metropolis-Hastings algorithm. We obtain $\mathbb{E}\gamma_{k}^{I}\mid \mathcal{D}^{I}$ and $\boldsymbol{\beta}_{k}^{I}\mid \mathcal{D}^{I}$ from the posterior samples after the algorithm converges.

\subsection{Empirical Null Based Variable Selection}
As the proposed posterior-informed shrinkage prior achieves shrinkage by the spike term, the variables can be selected by comparing posterior inclusion probability to a given cut-off value, which is $\widehat{\text{PIP}}_{k}^{I}>C_{k}^{I}$. In the external data, $C_{k}^{E}$ can be set as 0.5, which is a commonly used cut-off value in spike-and-slab prior \citep{MalsinerWalli2018}. An alternative cut-off value can be derived from Zcut statistics \citep{Ishwaran2005}. Considering the small sample size and different variable types, a general $C^{E}$ for every $X$ is not feasible. Hence, we adopted the empirical null calculated by permutation to obtain $\widehat{C}_{k}^{I}$, which avoids the abuse of external information and modeling the noise instead of the signal. The permutation-based empirical null is widely used in variable selection based on machine learning methods. For example, in variable selection based on Bayesian additive regression trees (BART), existing literature discussed extensively on BART growing trees with unrelated variables, which is taken as ``modeling the noise'', and empirical null can hugely improve variable selection accuracy \citep{Tan2019, Bleich2014}. With a combined $\mathbf{X}=\{\mathbf{X}^{E}, \mathbf{X}^{I}\}$ and $Y=\{Y^{E}, Y^{I}\}$, we resample $\tilde{Y}$ from $Y$ without replacement so that $\tilde{Y}$ is unrelated to $\mathbf{X}$ and the combined data has larger sample size to enable stable estimation of $\widehat{C}_{k}^{I}$. We estimate $\widehat{C}^{I}_{k}$ through bootstrap to reduce the uncertainty of $\widehat{C}^{I}_{k}$. Let $\mathcal{C}^{I}_{k}=\{\widehat{\text{PIP}}_{k,j}^{I}\}_{j=1}^{N}$ denote the estimated PIP of $\mathbf{X}_k$ under permutation, where $N$ is the count of total bootstrapped replicates. Then $\widehat{C}_{k}^{I}=\mathbb{E} \mathcal{C}^{I}_{k}$. The variables are then selected by $\mathbb{E}\gamma_{k}^{I}\mid \mathcal{D}^{I}>\widehat{C}^{I}_{k}$.

\subsection{Extension to Multiple External Data}
In basket trial designs, it is common to have multiple concurrent trials targeting similar disease subtypes or treatment regimens. Existing data borrowing methods often rely on pre-clustering data sources or assigning pre-computed weights to datasets based on manually defined similarity scores \citep{Ji2023, Jin2023}. In contrast, APSP adaptively weighs contributions from multiple external datasets and simultaneously estimates shrinkage parameters in a unified Bayesian framework.

To achieve this, we extend the original spike-and-slab structure by merging the spike-normal density components in Equation (\ref{eq:APSP}) and relaxing the borrowing mechanism for inclusion probabilities. Specifically, we generalize the model to accommodate multiple external datasets, denoted as $\mathcal{D}_{m}^{E} = \{Y_i, \mathbf{X}_i\}_{i=1}^{n_{m}^{E}}$, where $m \in \{1, \ldots, M\}$ indexes the external data sources. To simultaneously enable shrinkage and data borrowing, we employ a mixture informative shrinkage prior that combines three components: (i) a spike-normal density to induce sparsity, (ii) a non-informative normal density to allow unbiased estimation for weak signals, and (iii) an informative normal density derived from external data. The data source indicator is modeled using a multinomial distribution, governed by a Dirichlet prior that can be either informative or non-informative, depending on the level of prior knowledge available. This extended version of the APSP method preserves variable-specific borrowing and adapts dynamically across multiple external sources. The model can be formally expressed as:

\begin{longequation}
    \begin{array}{l}
        Y_{i}^{I} \sim \mathcal{N}(\mathbf{X}^{T}_{i}\boldsymbol{\beta}^{I}, \sigma_I^2),\ i=1,\cdots,n^{I}\\
        \beta_{k}^{I} \sim \nu_{k,1}\mathcal{N}(0, \tau_{k}^2)+\nu_{k,2}\mathcal{N}(0, m_{k}\tau_{k}^2) + \sum_{j=3}^{m+2}\nu_{k,j}\mathcal{N}(\widehat{\beta}_{k,j-2}, \tau_{k, j-2}^2\widehat{var}(\beta_{k,j-2}))\\
        \nu_{k} \sim Multinomial(\pi_{k,1},\cdots,\pi_{k,m+2}) \\
        \pi_{k,1},\cdots,\pi_{k,m+2} \sim Dirichlet(\alpha_{k,1},\cdots, \alpha_{k,m+2}) \\
        \alpha_{k,m} \sim Gamma(a_{\alpha_{k,m}}, b_{\alpha_{k,m}})\\
        \tau_{k, j-2}^2 \sim Gamma(\zeta_{k}, \xi_{k});
        \zeta_{k} \sim U(a_{\zeta}, b_{\zeta});
        \xi_{k} \sim U(a_{\xi}, b_{\xi})\\
        \sigma^2 \sim Gamma(a_{\sigma}, b_{\sigma}).
    \end{array}
    \label{eq:Dirchilet}
\end{longequation}

The hierarchical model described above retains the adaptive slab-normal component and introduces a latent variable $\nu_k$, which governs both the probability of borrowing information and the selection of the external data source from which to borrow. The multiple layers of latent variables facilitate fully data-driven information borrowing, allowing the algorithm to adaptively determine the extent of borrowing by comparing the similarity of $\beta_k$ estimates across multiple external sources and the internal (local) model. This structure enables variable-specific, source-specific borrowing decisions that are responsive to the observed data, thereby enhancing both robustness and estimation efficiency.

\section{Simulation}
We conducted extensive simulation studies to evaluate the robustness and performance of the proposed APSP method under a variety of scenarios that reflect different relationships between internal and external data signals. We considered the following four scenarios: Scenario 1 - Fully exchangeability, where the true regression coefficients $\boldsymbol{\beta}$ are identical between internal and external datasets. Scenario 2 - Both internal and external datasets share the same set of signal variables and noise variables, which means the external data and internal data have same signal pattern, but allowing for differences in effect sizes. Scenario 3 - External data and internal data have non-empty overlap of signals, which is the condition stated in Assumption 1. Scenario 4 - External data and internal data have no common signals, i.e., Assumption 1 is violated. In scenarios 3 and 4, variables that are informative in one dataset may be uninformative or even spurious in the other, and vice versa. These scenarios were designed to reflect a spectrum of real-world conditions, ranging from ideal compatibility to substantial heterogeneity between data sources.

To reflect the heterogeneity of covariate types commonly encountered in real-world studies, we simulated three types of covariates: binary variables, uniformly distributed variables, and normally distributed variables. The clinical endpoint variable is continuous and generated as a linear combination of all three types of covariates. The external dataset contains 50 observations, representing a relatively large sample size comparable to that of our motivating real-world dataset. To evaluate performance under small-sample conditions, we set the internal dataset sample size to 10, 20, and 30. These sample sizes are typical for ultra-rare disease or CGT trials. For each variable type, 2 out of 5 covariates are designated as signal variables, while the remaining 3 are treated as noise. The full data-generating process is also summarized in the Appendix A. 

We compared the proposed APSP with several popular and classic Bayesian variable selection and data borrowing methods, including the local spike-and-slab prior (SSP), LASSO, horseshoe prior (HP), power prior (PP), modified power prior (MPP), meta-analytic predictive prior (MAP) and commensurate prior (CP). All simulations were conducted with 500 replicates. 

APSP is designed to identify variables associated with the clinical endpoint efficiently and robustly. We evaluated variable selection performance with three metrics: (1) overall selection correctness, defined as the proportion of variables correctly selected or excluded; (2) false discovery rate (FDR), quantifying the proportion of selected variables that are noise; and (3) true discovery rate (TDR), measuring the proportion of signal variables correctly identified. In addition, we assessed parameter estimation accuracy by computing the root mean square error (RMSE) of the estimated regression coefficients $\widehat{\boldsymbol{\beta}}$. 

\subsection{Simulation Results}
Table~\ref{table:table1} summarizes the performance of all methods in terms of variable selection accuracy. APSP performs best when the external and internal data share no common signals across a range of internal sample sizes, while performing comparably to the best methods in the other scenarios. In Scenarios~1 and~2, the signal patterns are the same in the external and internal data, which favors direct data-borrowing methods such as PP and MPP; the more adaptive approaches, such as CP and MAP, show poorer performance because some internal information is used to assess similarity between the external and internal data. As a two-step fitting procedure, APSP also partially relies on accurate modeling of the external data, similar to MAP. Local shrinkage methods, such as LASSO and HP, achieve relatively satisfactory variable selection accuracy when the internal sample size is 30; as expected, their performance decreases as the internal sample size decreases. 

\begin{table}
    \caption{Simulation result for overall selection correctness from Scenario 1 to Scenario 4, with varying internal data sample size. The selection correctness is reported as the percentage of correctly decided variables ($\%\ (sd)$). High overall selection correctness is favored. 
    }
    \label{tbl:table2d1s}
    \centering
    \resizebox{\textwidth}{!}{%
\begin{tabular}{ccccccccc}
\hline
\hline
\multicolumn{9}{c}{\textbf{Internal Data Sample Size: 30}}\\
\hline
\multicolumn{1}{c}{} & 
\multicolumn{1}{c}{APSP} 
&\multicolumn{1}{c}{SSP} 
& \multicolumn{1}{c}{LASSO} 
& \multicolumn{1}{c}{HP}
& \multicolumn{1}{c}{PP}
& \multicolumn{1}{c}{MAP}
& \multicolumn{1}{c}{MPP}
& \multicolumn{1}{c}{CP}\\
\hline
    Identical Signals & 83.1 (8.6) & 63.0 (4.7) & 64.4 (5.7) & 80.1 (5.9) & 90.4 (6.2) & 79.1 (6.3) & \textbf{90.5 (6.2)} & 84.1 (6.9) \\
    Same Pattern & 86.0 (8.1) & 64.5 (6.0) & 69.4 (8.9) & 86.0 (5.1) & 91.2 (6.0) & 85.5 (5.0) & \textbf{91.3 (6.0)} & 87.9 (5.7) \\
    Partial Overlap & 86.8 (9.1) & 62.1 (4.3) & 63.1 (6.3) & 86.4 (6.6) & 86.1 (8.8) & 84.0 (8.6) & 84.1 (8.7) & \textbf{92.0 (6.7)} \\
    No Overlap & \textbf{87.8 (9.1)} & 61.3 (3.4) & 63.1 (6.4) & 86.0 (7.3) & 58.9 (12.2) & 81.9 (9.0) & 55.9 (12.6) & 86.6 (8.1) \\
\hline
\hline
\multicolumn{9}{c}{\textbf{Internal Data Sample Size: 20}}\\
\hline
\multicolumn{1}{c}{} & 
\multicolumn{1}{c}{APSP} 
&\multicolumn{1}{c}{SSP} 
& \multicolumn{1}{c}{LASSO} 
& \multicolumn{1}{c}{HP}
& \multicolumn{1}{c}{PP}
& \multicolumn{1}{c}{MAP}
& \multicolumn{1}{c}{MPP}
& \multicolumn{1}{c}{CP}\\
\hline
    Identical Signals & 77.5 (9.8) & 60.8 (2.4) & 60.0 (0.0) & 72.8 (6.1) & \textbf{89.2 (6.6)} & 72.9 (6.7) & 89.2 (6.6) & 80.2 (6.1) \\
    Same Pattern & 79.9 (10.5) & 60.9 (2.5) & 60.1 (0.7) & 78.6 (6.5) & \textbf{89.5 (6.2)} & 78.7 (7.6) & 89.5 (6.1) & 84.4 (5.7) \\
    Partial Overlap & 79.7 (11.2) & 60.3 (1.5) & 60.1 (0.6) & 75.5 (7.0) & 79.0 (9.1) & 70.1 (8.4) & 78.0 (8.9) & \textbf{82.3 (8.6)} \\
    No Overlap & \textbf{79.7 (11.6)} & 60.3 (1.6) & 60.0 (0.0) & 74.7 (8.2) & 48.9 (10.4) & 68.1 (8.1) & 47.1 (10.6) & 73.3 (9.2) \\
\hline
\hline
\multicolumn{9}{c}{\textbf{Internal Data Sample Size: 10}}\\
\hline
\multicolumn{1}{c}{} & 
\multicolumn{1}{c}{APSP} 
&\multicolumn{1}{c}{SSP} 
& \multicolumn{1}{c}{LASSO} 
& \multicolumn{1}{c}{HP}
& \multicolumn{1}{c}{PP}
& \multicolumn{1}{c}{MAP}
& \multicolumn{1}{c}{MPP}
& \multicolumn{1}{c}{CP}\\
\hline
    Identical Signals & 72.1 (8.9) & 60.1 (0.9) & 60.0 (0.0) & 63.4 (3.8) & \textbf{88.3 (6.4)} & 63.6 (4.0) & 88.3 (6.3) & 67.9 (5.1) \\
    Same Pattern & 72.2 (9.7) & 60.1 (0.9) & 60.0 (0.0) & 64.4 (4.4) & \textbf{88.6 (6.6)} & 64.7 (4.8) & 88.6 (6.7) & 70.6 (5.6) \\
    Partial Overlap & 69.4 (10.5) & 60.0 (0.4) & 60.0 (0.0) & 62.6 (4.0) & 72.6 (7.2) & 61.1 (2.8) & \textbf{72.7 (7.1)} & 64.7 (5.1) \\
    No Overlap & \textbf{65.9 (10.4)} & 60.0 (0.4) & 60.0 (0.0) & 62.6 (3.7) & 39.9 (9.3) & 60.6 (2.2) & 40.0 (9.4) & 60.6 (3.8) \\
\hline
\end{tabular}%
}\label{table:table1}
\end{table}

\begin{figure}
    \centering
    \includegraphics[width=1\linewidth]{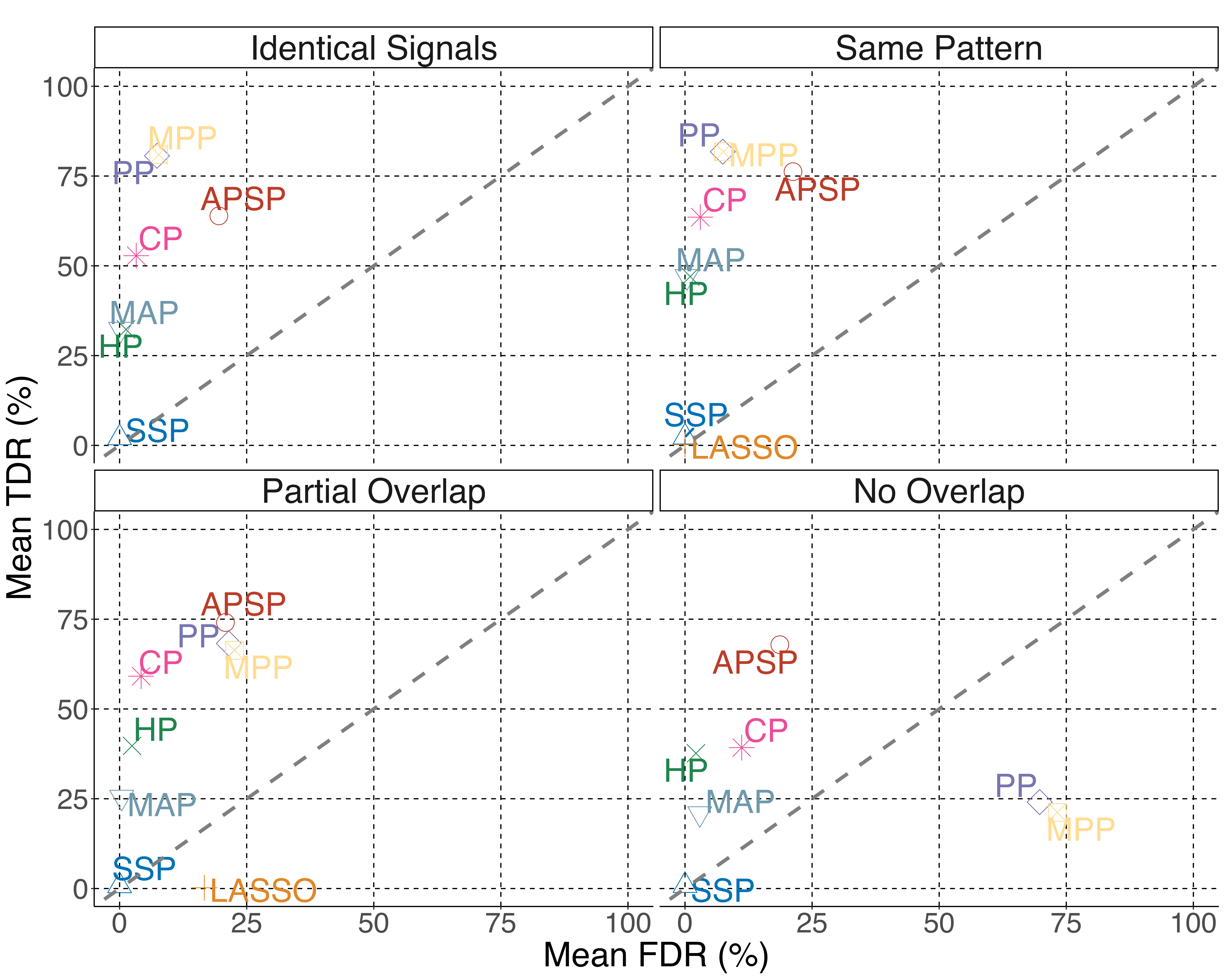}
    \caption{Scatter plot of TDR and FDR of proposed method and competing methods under multiple scenarios with internal data sample size at 20. Smaller RMSE means better estimation of model parameters. High TDR and low FDR are favored, thus the methods locates at top-left is better.
    }
    \label{fig:fig1}
\end{figure}

Figure~\ref{fig:fig1} further demonstrates each method’s performance in terms of FDR and TDR, acknowledging that researchers may prioritize detecting true signals or avoiding the selection of spurious variables depending on their goals. Consistent with Table~\ref{table:table1}, APSP attains the best or near-best mean TDR in Scenarios~1–3, with acceptable mean FDR. In Scenario~4, APSP is substantially more efficient than local shrinkage methods in mean TDR and attains much lower mean FDR than direct data-borrowing approaches.

\vspace{-1pt}
    
In addition, Figure~\ref{fig:fig2} shows the RMSE of \(\widehat{\boldsymbol{\beta}}\) across all methods with an internal sample size of $n^I=20$ and we summarized RMSE results with th other internal sample sizes of $n^I=10/30$ in Appendix B. For all methods, we estimate \(\boldsymbol{\beta}^{I}\) using the posterior mean. Consistent with what we observe for variable selection, APSP exhibits slightly higher RMSE than PP and MPP when full data pooling is the most efficient approach. APSP has comparable performance to adaptive methods such as MAP and CP, and better performance than local-shrinkage methods, which were not primarily designed for parameter estimation. APSP achieves this performance because, when information borrowing is plausible, it uses both the external and internal data to obtain the posterior distribution; when no information can be borrowed, the shrinkage component in APSP strongly shrinks posterior means toward zero.

\vspace{-1pt}
\begin{figure}
    \centering
    \includegraphics[width=1\linewidth]{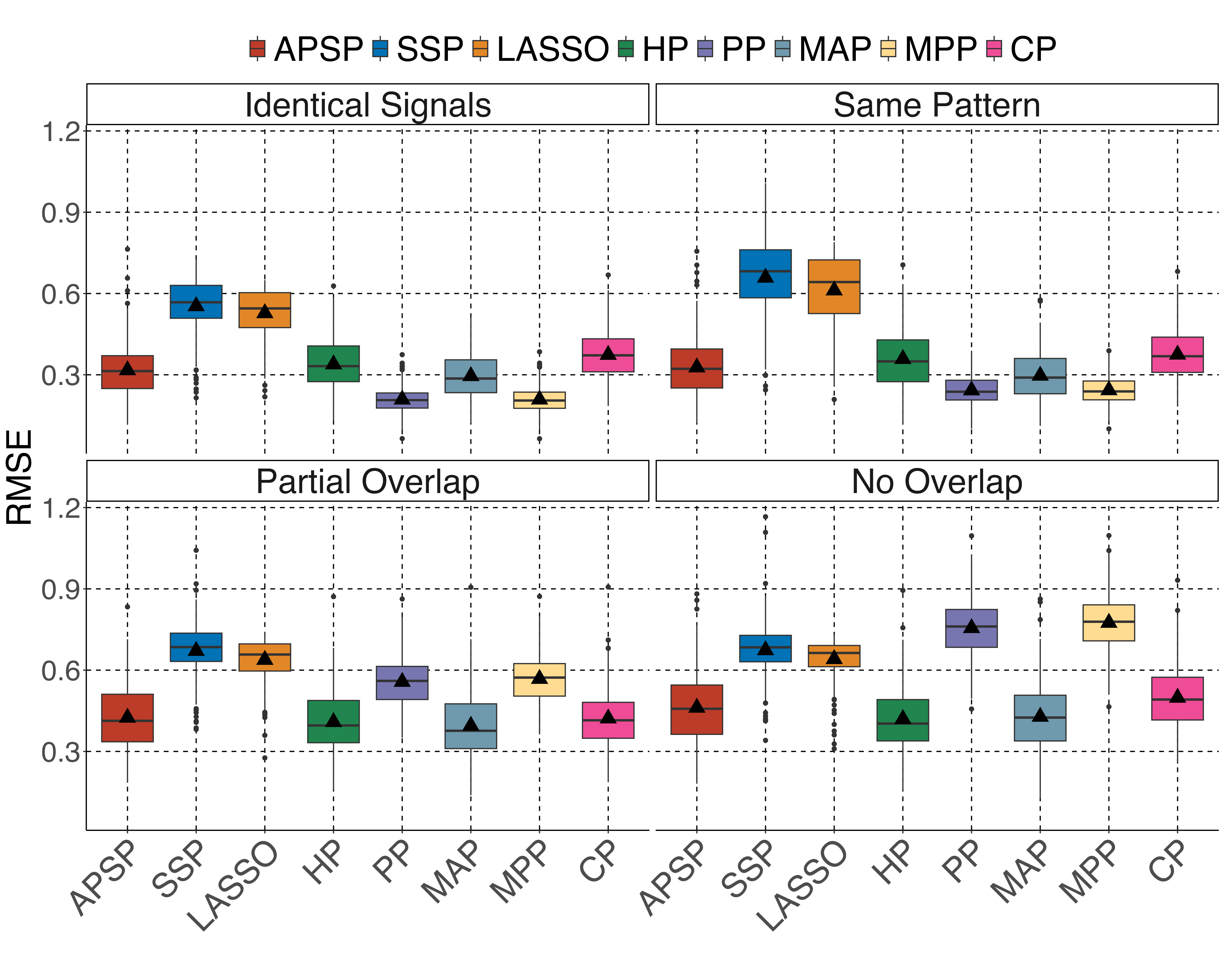}
    \caption{Root Mean Square Error (RMSE) of proposed method and competing methods under multiple scenarios with internal data sample size at 20. Smaller RMSE means better estimation of model parameters.
    }
    \label{fig:fig2}
\end{figure}

\subsection{Simulation Conclusions}
Overall, APSP is an efficient and robust method for variable selection with small samples. In addition, the posterior mean can yield biased yet useful estimates of signal effect sizes. This efficiency and robustness provide the foundation for applying the proposed method to real-data applications where the ground truth is unknown. Some existing methods may perform well when certain assumptions are met, but their performance degrades substantially when those assumptions are violated. This limits the large-scale application of such methods, as the assumptions are difficult or even impossible to verify in real data.

\section{Application to Type 1 Diabetes Data}
Type 1 diabetes (T1D) affects millions of people in the United States, and often develops in children and young adults \citep{Holt2021Type1Consensus}. As opposed to type 2 diabetes, T1D is characterized primarily by the patient's own immune system attacking the Beta-islet cells in the body, thereby causing the patient to be unable to produce insulin and hence, unable to regulate their blood glucose. The standard therapy involves the administration of exogenous insulin, delivered either through multiple daily injections or via continuous infusion using a wearable insulin pump. However, insulin therapies often fail to achieve normal glycemic control, and long-term T1D survivors frequently suffer from vascular complications such as diabetic retinopathy and diabetic nephropathy. To explore alternative treatment strategies and benefit subjects with severe T1D, the Clinical Islet Transplantation Consortium (CITC) was established to conduct and coordinate research on islet transplantation therapies. Among the series of clinical islet transplantation (CIT) trials led by the consortium, we focus on two studies with closely related designs: CIT06 and CIT07. CIT07 is a prospective, single-arm, multi-center trial with sample size $N=48$, evaluating the efficacy of islet transplantation in the general T1D population \citep{Markmann2021}. In contrast, CIT06 is a prospective, single-arm, multi-center trial with much smaller sample size $N=24$, assessing the efficacy of islet transplantation in T1D patients who have previously received kidney transplants \citep{Hering2016}.

\subsection{Data Preparation}
In the CIT06 trial, due to the strict inclusion criteria limiting participants to T1D patients with prior kidney transplantation, the sample size is 24. This small sample size presents significant challenges for identifying variables associated with the clinical endpoints of interest. In contrast, the CIT07 trial used the same therapeutic protocol but enrolled a broader population of T1D patients with severe hypoglycemic events, resulting in a slightly larger sample size of 48. The shared treatment protocol form a strong basis for borrowing information across the two datasets. At the same time, the biological and clinical differences between the two populations highlight the necessity of adaptive information borrowing to account for potential discrepancies in the associations between the covariates and the clinical endpoint of interest.

In our analysis, CIT07 is treated as the external dataset, while CIT06 is the internal dataset of primary interest. Our goal is to identify variables associated with peak c-peptide measured from the mixed meal tolerance test (MMTT) in CIT06. Peak c-peptide is of interest because c-peptide is a byproduct when the body produces insulin \citep{Venugopal2023StatPearls,Latres2024}. Thus, higher c-peptide values can be interpreted as robust insulin production by the body when stimulated by glucose indicating healthy Beta-islet cell engraftment \citep{Forbes2018MMTT,Hering2016CIT07,Latres2024}.
Descriptive statistics for both datasets are summarized in Table \ref{tbl:table2}. Variables included for selection are based on clinically relevance and missingness: Age, T1D duration (yrs), race (white/non-white), gender, BMI, insulin use (insulin/kg), HbA1c at baseline, amount of transplanted islet cells (IEQ/kg). Additionally, the race variable was collapsed into a binary indicator (white vs. non-white), as the limited sample size does not support subgroup analyses for multiple racial categories. We log-transformed amount of transplanted islet cells to the recipient and by protocol, the amount of transplanted islet cell should be at least 5000. Some recipients received higher dose, which makes the distribution right skewed. Our clinical endpoint of interest is set as peak c-peptide at Day 75, derived by maximum of fasting c-peptide, stimulated c-peptide at 60 min and stimulated c-peptide at 90 min at Day 75 with respect to the first dose each patient receives. For patients who dropped out or had their second dose before Day 75, we imputed their peak c-peptide as $0.05$ ng/mL, which is the lower bound of detection in the trial with the assumption of insufficient insulin production after the initial dose. For subjects who missed their Day 75 visit, the closet measurement (before the second dose where applicable) was used to impute the peak c-peptide at Day 75. If this measure was more than or equal to 2 days apart from the Day 75 visit window, no imputation was made. A sensitivity analysis was conducted with complete cases only.  
\begin{table}
    \caption{Sumamry statistics of baseline variables in CIT06 and CIT07. Continuous variables are described as Mean (sd) and (Min, Max), and categorical variables are described as $n\ (\%)$. Missingness of each variable within each data is summarized as $n\ (\%)$.
    }
    \label{tbl:table2}
    \centering
    \resizebox{\textwidth}{!}{%
\begin{tabular}{lcccccc}
\hline
\hline
     & & CIT06 (N=24) & & & CIT07 (N=48) & \\
\hline
    Variable & \textbf{Mean (sd)} & \textbf{(Min, Max)} & \textbf{Missing n (\%)}  & \textbf{Mean (sd)} & \textbf{(Min, Max)} & \textbf{Missing n (\%)} \\
\hline
    Age (yrs) & 51.8 (11.1) & (29.2, 69.6) & 0 (0) & 47.8 (11.5) & (26.2, 65.5) & 0 (0) \\
    T1D Duration (yrs) & 37.0 (10.0) & (17.0, 55.0) & 0 (0) & 31.5 (11.0) & (11.0, 57.0) & 0 (0) \\
    BMI & 24.6 (3.1) & (18.9, 30.4) & 0 (0) & 24.9 (3.1) & (18.9, 29.8) & 0 (0) \\
    Gender & & & & & & \\
    \ \ Male & 13 (54.1\%) &  & 0 (0) & 19 (39.6\%) &  & 0 (0) \\
    Race & & & & & & \\
    \ \ White & 6 (25\%) &  & 0 (0) & 6 (12.5\%) &  & 0 (0) \\
    log(IEQ/kg) & 8.9 (0.3) & (8.5, 9.5) & 0 (0.0) & 8.9 (0.2) & (8.5, 9.4) & 0 (0) \\
    Insulin/kg & 0.5 (0.1) & (0.3, 0.7) & 1 (4.2) & 0.5 (0.1) & (0.2, 0.8) & 0 (0) \\
    HbA1c (\%) & 8.2 (1.5) & (6.0, 12.7) & 0 (0) & 7.4 (0.9) & (5.7, 9.2) & 0 (0) \\
\hline
\hline
\end{tabular}%
}
\end{table}

\subsection{Identification of Variables Associated with Peak C-peptide at Day 75}
\begin{figure}
    \centering
    \includegraphics[width=1\linewidth]{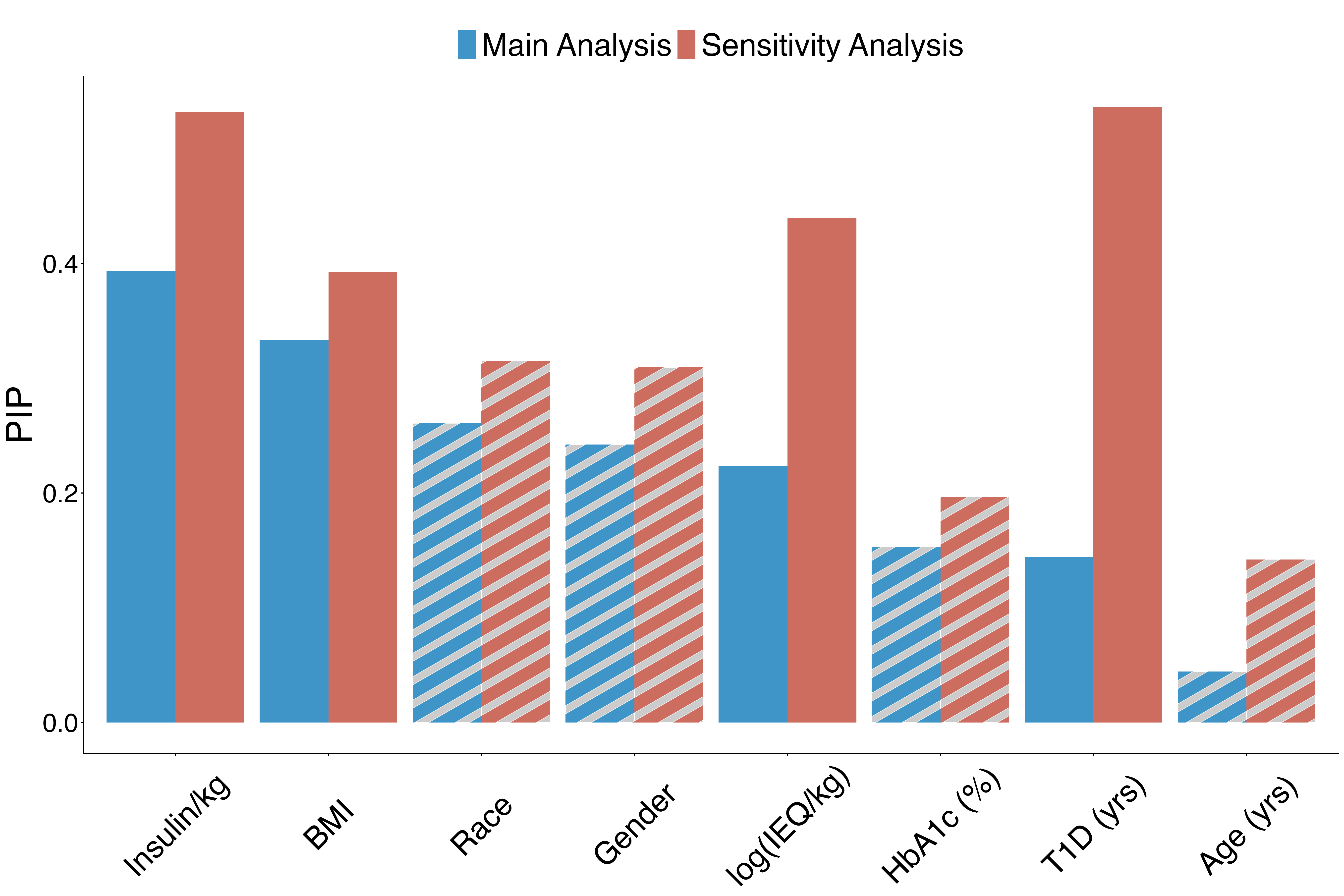}
    \caption{Posterior inclusion probabilities (PIP) of variables under the APSP prior in the main and sensitivity analyses. Filled bars indicate variables selected, while striped bars indicate variables excluded based on the empirical null criterion. Variables are ranked by PIP from highest to lowest in the main analysis, displayed left to right.}
    \label{fig:fig3}
\end{figure}

The number of subjects included in our main analysis is 48 for CIT07 and 23 for CIT06. One patient was missing their insulin use at baseline for CIT06. For the sensitivity analysis, the number of subjects was 46 and 18 respectively. Classical variable selection methods such as spike-and-slab prior, Horseshoe prior, LASSO and data borrowing methods including modified power prior and commensurate prior were applied on CIT06 data as reference. Among the reference methods, spike-and-slab prior model selected no variables due to the small sample size and modified power prior model only selected insulin use. LASSO, horseshoe prior model and commensurate prior model selected different variables in the main analysis and sensitivity analysis. APSP yielded consistent results of selected variables in main analysis and sensitivity analysis as shown in Table~\ref{tbl:table3} and Figure~\ref{fig:fig3}. APSP reported that BMI, insulin use, duration of T1D, and the amount of transplanted islet cells were associated with peak c-peptide at Day 75. 

Association of BMI and insulin use with peak c-peptide aligns with our medical knowledge \citep{Forbes2016BETA2, Lam2022BETA2Early}. Insulin use is a predictor of of graft decline and loss of insulin independence after pancreatic islet allotransplantation at Day 75 \citep{Lam2022BETA2}. Higher amounts of transplanted islet cells, indicates that having more transplanted cells will enable the recipient to have better ability of producing insulin and thus have higher peak c-peptide. Duration of T1D reflects the patient's disease status with the hypothesis that if a patient is more diseased, it would trigger a larger production of insulin and hence, c-peptide. Currently this hypothesis needs more clinical evidence to support. 

\begin{table}
    \caption{Summary of $\widehat{\boldsymbol{\beta}}$ and its $95\%$ credible interval. Bold text denotes significance under $95\%$ credible interval not across 0 (HP, MPP, CP) or $PIP>\widehat{C}^{I}$ (APSP) or $PIP>0.5$ (SSP). $\widehat{C}^{I}$ is a permutation empirical null.
    }
    \label{tbl:table3}
    \centering
    \resizebox{\textwidth}{!}{%
\begin{tabular}{lcccccc}
\hline
\hline
    \multicolumn{7}{c}{Main Analysis}\\
\hline
    Variable & $\boldsymbol{\beta}^{APSP}\ (95\% CI)$ & $\boldsymbol{\beta}^{SSP}\ (95\% CI)$ & $\boldsymbol{\beta}^{LASSO}$ & $\boldsymbol{\beta}^{HP}\ (95\% CI)$ & $\boldsymbol{\beta}^{MPP}\ (95\% CI)$ & $\boldsymbol{\beta}^{CP}\ (95\% CI)$ \\
\hline
    Age (yrs) & 0.01 (-0.04, 0.07) & 0.00 (0.00, 0.00) & 0.00 & -0.02 (-0.08, 0.02) & 0.02 (-0.03, 0.07) & 0.01 (-0.09, 0.11)\\
    T1D (yrs) & \textbf{0.04 (-0.02, 0.14)} & 0.01 (0.00, 0.13) & \textbf{0.05} & \textbf{0.09 (0.02, 0.16)} & 0.01 (-0.04, 0.06) & 0.11 (0.00, 0.21)\\
    BMI & \textbf{0.09 (-0.04, 0.40)} & 0.01 (0.00, 0.14) & \textbf{0.09} & 0.02 (-0.04, 0.10) & 0.09 (-0.04, 0.22) & \textbf{0.27 (0.02, 0.52)} \\
    Gender & 0.00 (-0.81, 0.82) & 0.06 (-0.07, 1.14) & 0.00 & 0.02 (-0.43, 0.54) & -0.18 (-1.09, 0.73) & 0.16 (-1.18, 1.61) \\
    Race & 0.05 (-0.67, 1.03) & 0.02 (-0.32, 0.71) & 0.00 & 0.03 (-0.4, 0.58) & 0.03 (-1.00, 1.06) & 0.06 (-1.24, 1.39) \\
    log(IEQ/kg) & \textbf{0.07 (-0.13, 0.78)} & 0.01 (0.00, 0.07) & 0.00 & 0.06 (-0.18, 0.40) & 0.28 (-0.12, 0.68) & 0.70 (-0.01, 1.41) \\
    Insulin/kg & \textbf{0.76 (-0.07, 3.62)} & 0.54 (-2.18, 6.10) & 0.00 & -0.03 (-1.07, 0.84) & \textbf{3.31 (0.27, 6.4)} & 2.52 (-0.13, 6.54)\\
    HbA1c (\%) & 0.03 (-0.08, 0.44) & 0.00 (0.00, 0.00) & 0.00 & 0.01 (-0.15, 0.18) & 0.09 (-0.25, 0.42) & 0.48 (-0.06, 0.98) \\
\hline
    \multicolumn{7}{c}{Sensitivity Analysis}\\
\hline
    Variable & $\boldsymbol{\beta}^{APSP}\ (95\% CI)$ & $\boldsymbol{\beta}^{SSP}\ (95\% CI)$ & $\boldsymbol{\beta}^{LASSO}$ & $\boldsymbol{\beta}^{HP}\ (95\% CI)$ & $\boldsymbol{\beta}^{MPP}\ (95\% CI)$ & $\boldsymbol{\beta}^{CP}\ (95\% CI)$ \\
\hline
    Age (yrs) & -0.02 (-0.17, 0.06) & 0.00 (0.00, 0.06) & 0.00 & \textbf{-0.12 (-0.21, -0.01)} & 0.02 (-0.04, 0.07) & -0.09 (-0.24, 0.05) \\
    T1D (yrs) & \textbf{0.11 (-0.01, 0.31)} & 0.03 (0.00, 0.18) & \textbf{0.10} & \textbf{0.23 (0.11, 0.35)} & 0.01 (-0.03, 0.06) & \textbf{0.23 (0.08, 0.38)}\\
    BMI & \textbf{0.10 (-0.04, 0.41)} & 0.03 (0.00, 0.37) & \textbf{0.17} & 0.05 (-0.04, 0.17) & 0.01 (-0.03, 0.06) & 0.22 (-0.01, 0.44) \\
    Gender & -0.05 (-1.10, 0.77) & 0.04 (-0.56, 1.26) & 0.00 & -0.04 (-0.77, 0.59) & -0.22 (-1.12, 0.68) & -0.19 (-1.53, 1.24) \\
    Race & 0.00 (-0.96, 0.98) & 0.01 (-0.64, 0.89) & 0.00 & -0.1 (-1.02, 0.46) & -0.09 (-1.13, 0.95) & -0.27 (-1.66, 1.09) \\
    log(IEQ/kg) & \textbf{0.27 (-0.08, 1.23)} & 0.05 (0.00, 0.87) & \textbf{0.37} & 0.53 (-0.03, 1.22) & 0.31 (-0.09, 0.72) & \textbf{0.92 (0.17, 1.67)} \\
    Insulin/kg & \textbf{0.85 (-0.29, 3.98)} & 1.91 (-1.03, 9.36) & \textbf{2.21} & 0.14 (-0.94, 2.00) & \textbf{4.19 (1.1, 7.21)} & 1.97 (-0.79, 6.99) \\
    HbA1c (\%) & -0.03 (-0.48, 0.21) & -0.01 (-0.05, 0.00) & 0.00 & -0.15 (-0.53, 0.08) & 0.04 (-0.34, 0.43) & -0.01 (-0.60, 0.60) \\
\hline
\hline
\end{tabular}%
}
\end{table}

\vspace{-1pt}

\section{Conclusion and Discussion}
In this work, we proposed APSP, a robust and flexible Bayesian method to borrow information from external data to facilitate better variable selection on small sample trial data. APSP takes advantage of the form of spike-and-slab prior, which enables shrinking small effects towards zero and better variable selection accuracy when facing wide credible intervals. Compared to the existing data borrowing methods, APSP is flexible to deal with possible discrepancies between the external data and the internal data. Simulation results shows that APSP has robust performance against multiple external data and internal data relationships and internal data sample sizes. Application of empirical null with permutation yields more robust and flexible cut-off values under small sample sizes, where the external information may dominant internal data and the model may be overfitting when the signal is weak.

Finally, some extension can be made based on the form of APSP, including extension to multiple external data sources and other outcome types such as binary endpoint. With the development of shrinkage priors with other structures and distributions, APSP can also takes other prior structures with more efficiency of shrinkage.   

\label{lastpage}
\backmatter
\bibliographystyle{biom} 
\bibliography{Reference}

@article{Bleich2014,
  title={Variable selection for BART: An application to gene regulation},
  author={Bleich, Justin and Kapelner, Adam and George, Edward I. and Jensen, Shane T.},
  journal={Annals of Applied Statistics},
  volume={8},
  number={3},
  pages={1750--1781},
  year={2014},
  month={September},
  doi={10.1214/14-AOAS755},
  url={https://doi.org/10.1214/14-AOAS755}
}

@article{Carvalho2010,
  title={The horseshoe estimator for sparse signals},
  author={Carvalho, C. M. and Polson, N. G. and Scott, J. G.},
  journal={Biometrika},
  volume={97},
  number={2},
  pages={465--480},
  year={2010},
  url={http://www.jstor.org/stable/25734098}
}

@article{Freedman1983,
  title={A note on screening regression equations},
  author={Freedman, D. A.},
  journal={The American Statistician},
  volume={37},
  number={2},
  pages={152--155},
  year={1983},
  doi={10.2307/2685877},
  url={https://doi.org/10.2307/2685877}
}

@article{Hans2009,
  title={Bayesian lasso regression},
  author={Hans, C.},
  journal={Biometrika},
  volume={96},
  number={4},
  pages={835--845},
  year={2009}
}

@book{Hastie2001,
  title={The Elements of Statistical Learning},
  author={Hastie, Trevor and Tibshirani, Robert and Friedman, Jerome},
  year={2001},
  publisher={Springer},
  address={New York, NY, USA}
}

@article{Hering2016,
  title={Phase 3 trial of transplantation of human islets in type 1 diabetes complicated by severe hypoglycemia},
  author={Hering, B. J. and Clarke, W. R. and Bridges, N. D. and Eggerman, T. L. and Alejandro, R. and Bellin, M. D. and others},
  journal={Diabetes care},
  volume={39},
  number={7},
  pages={1230--1240},
  year={2016}
}

@article{Hobbs2011,
  title={Hierarchical commensurate and power prior models for adaptive incorporation of historical information in clinical trials},
  author={Hobbs, B. P. and  Carlin, B. P. and Mandrekar, S. J. and Sargent, D. J.},
  journal={Biometrics},
  volume={67},
  number={3},
  pages={1047--1056},
  year={2011}
}

@article{Hobbs2012,
  title={Commensurate priors for incorporating historical information in clinical trials using general and generalized linear models},
  author={Hobbs, B. P. and Sargent, D. J. and Carlin, B. P.},
  journal={Bayesian Analysis},
  volume={7},
  number={3},
  pages={639--674},
  year={2012}
}

@article{Ibrahim2015,
  title={The power prior: theory and applications},
  author={Ibrahim, J. G. and Chen, M. H. and Gwon, Y. and Chen, F.},
  journal={Statistics in Medicine},
  volume={34},
  number={28},
  pages={3724--3749},
  year={2015}
}

@article{Ishwaran2005,
  title={Spike and slab variable selection: frequentist and Bayesian strategies},
  author={Ishwaran, H. and Rao, J. S.},
  journal={The Annals of Statistics},
  volume={33},
  number={2},
  pages={730--773},
  year={2005}
}

@article{Ji2023,
  title={A flexible Bayesian framework for individualized inference via adaptive borrowing},
  author={Ji, Z. and Wolfson, J.},
  journal={Biostatistics},
  volume={24},
  number={3},
  pages={669--685},
  year={2023}
}

@article{Jiang2023,
  title={Elastic priors to dynamically borrow information from historical data in clinical trials},
  author={Jiang, L. and Nie, L. and Yuan, Y.},
  journal={Biometrics},
  volume={79},
  number={1},
  pages={49--60},
  year={2023}
}

@article{Jin2023,
  title={Bayesian adaptive design for covariate‐adaptive historical control information borrowing},
  author={Jin, H. and Kim, M. O. and Scheffler, A. and Jiang, F.},
  journal={Statistics in Medicine},
  volume={42},
  number={29},
  pages={5338--5352},
  year={2023}
}

@article{Markmann2021,
  title={Phase 3 trial of human islet‐after‐kidney transplantation in type 1 diabetes},
  author={Markmann, J. F. and Rickels, M. R. and Eggerman, T. L. and Bridges, N. D. and Lafontant, D. E. and Qidwai, J. and others},
  journal={American Journal of Transplantation},
  volume={21},
  number={4},
  pages={1477--1492},
  year={2021}
}

@article{MalsinerWalli2018,
  title={Comparing spike and slab priors for Bayesian variable selection},
  author={Malsiner-Walli, Gertraud and Wagner, Helga},
  journal={arXiv preprint arXiv:1812.07259},
  year={2018}
}

@article{Ouma2022,
  title={Bayesian modelling strategies for borrowing of information in randomised basket trials},
  author={Ouma, L. O. and Grayling, M. J. and Wason, J. M. and Zheng, H.},
  journal={Journal of the Royal Statistical Society: Series C (Applied Statistics)},
  volume={71},
  number={5},
  pages={2014--2037},
  year={2022}
}

@article{Schmidli2014,
  title={Robust meta-analytic-predictive priors in clinical trials with historical control information},
  author={Schmidli, Heinz and Gsteiger, Sandro and Roychoudhury, Satrajit and O'Hagan, Anthony and Spiegelhalter, David and Neuenschwander, Beat},
  journal={Biometrics},
  volume={70},
  number={4},
  pages={1023--1032},
  year={2014},
  month={December},
  doi={10.1111/biom.12242},
  url={https://doi.org/10.1111/biom.12242}
}

@article{Tan2019,
  title={Bayesian additive regression trees and the General BART model},
  author={Tan, Y. V. and Roy, J.},
  journal={Statistics in Medicine},
  volume={38},
  number={25},
  pages={5048--5069},
  year={2019}
}

@article{vandeGeer2011,
  title={The adaptive and the thresholded Lasso for potentially misspecified models (and a lower bound for the Lasso)},
  author={van de Geer, Sara and B{\"u}hlmann, Peter and Zhou, Shuheng},
  journal={Electronic Journal of Statistics},
  volume={5},
  pages={688--749},
  year={2011},
  doi={10.1214/11-EJS624},
  url={https://doi.org/10.1214/11-EJS624}
}

@article{Yang2023,
  title={SAM: Self-Adapting Mixture Prior to Dynamically Borrow Information from Historical Data in Clinical Trials},
  author={Yang, P. and Zhao, Y. and Nie, L. and Vallejo, J. and Yuan, Y.},
  journal={Biometrics},
  volume={79},
  number={4},
  pages={2857--2868},
  year={2023}
}

@article{Zheng2022,
  title={Borrowing of information across patient subgroups in a basket trial based on distributional discrepancy},
  author={Zheng, H. and Wason, J. M.},
  journal={Biostatistics},
  volume={23},
  number={1},
  pages={120--135},
  year={2022}
}

@article{Zheng2023,
  title={Bayesian sample size determination using commensurate priors to leverage preexperimental data},
  author={Zheng, H. and Jaki, T. and Wason, J. M.},
  journal={Biometrics},
  volume={79},
  number={2},
  pages={669--683},
  year={2023}
}

@article{Zhou2024,
  title={Bayesian methods for information borrowing in basket trials: An overview},
  author={Zhou, T. and Ji, Y.},
  journal={Cancers},
  volume={16},
  number={2},
  pages={251},
  year={2024}
}

@article{Latres2024,
  title={Evidence for C-Peptide as a Validated Surrogate to Predict Clinical Benefits in Trials of Disease-Modifying Therapies for Type 1 Diabetes},
  author={Latres, Eduardo and Sweet, Ian R. and Rickels, Michael R. and Narendran, Parth and Skyler, Jay S. and Lachin, John M. and Jones, Andrew G. and Greenbaum, Carla J. and Sosenko, Jay M. and the JDRF-COVAC/C-Path–TrialNet T1D C-Peptide Biomarker Group},
  journal={Diabetes},
  volume={73},
  number={6},
  pages={823--833},
  year={2024},
  doi={10.2337/dbi23-0012},
  url={https://doi.org/10.2337/dbi23-0012}
}

@article{Forbes2018MMTT,
  title={Comparison of metabolic responses to the mixed meal tolerance test vs the oral glucose tolerance test after successful clinical islet transplantation},
  author={Forbes, Stephen and Lam, Anthony and Koh, Alexander and Imes, Sara and Dinyari, Parastoo and Malcolm, Andrew J. and Shapiro, A. M. James and Senior, Peter A.},
  journal={Clinical Transplantation},
  volume={32},
  number={8},
  pages={e13301},
  year={2018},
  doi={10.1111/ctr.13301},
  url={https://doi.org/10.1111/ctr.13301}
}

@article{Hering2016CIT07,
  title={Phase 3 Trial of Transplantation of Human Islets in Type 1 Diabetes Complicated by Severe Hypoglycemia},
  author={Hering, Bernhard J. and Clarke, William R. and Bridges, Nancy D. and Eggerman, Thomas L. and Alejandro, Rodolfo and Bellin, Melena D. and Chaloner, Kathryn and Czarniecki, Christine W. and Goldstein, Julia S. and Hunsicker, Lawrence G. and Kaufman, Dixon B. and Korsgren, Olle and Larsen, Christian P. and Luo, Xunrong and Markmann, James F. and Naji, Ali and Oberholzer, Jose and Posselt, Andrew M. and Rickels, Michael R. and Ricordi, Camillo and Robien, Mark A. and Senior, Peter A. and Shapiro, A. M. James and Stock, Peter G. and Turgeon, Nicole A.; Clinical Islet Transplantation Consortium},
  journal={Diabetes Care},
  volume={39},
  number={7},
  pages={1230--1240},
  year={2016},
  doi={10.2337/dc15-1988},
  url={https://doi.org/10.2337/dc15-1988}
}

@article{Forbes2016BETA2,
  title={Validation of the BETA-2 Score: An Improved Tool to Estimate Beta Cell Function After Clinical Islet Transplantation Using a Single Fasting Blood Sample},
  author={Forbes, Stephen and Oram, Richard A. and Smith, Andrew and Lam, Anthony and Olateju, Tinu and Imes, Sara and Malcolm, Andrew J. and Shapiro, A. M. James and Senior, Peter A.},
  journal={American Journal of Transplantation},
  volume={16},
  number={9},
  pages={2704--2713},
  year={2016},
  doi={10.1111/ajt.13807},
  url={https://doi.org/10.1111/ajt.13807}
}

@article{Lam2022BETA2Early,
  title={Estimation of Early Graft Function Using the BETA-2 Score Following Clinical Islet Transplantation},
  author={Lam, Anthony and Oram, Richard A. and Forbes, Stephen and Olateju, Tinu and Malcolm, Andrew J. and Imes, Sara and Shapiro, A. M. James and Senior, Peter A.},
  journal={Transplant International},
  volume={35},
  pages={10335},
  year={2022},
  doi={10.3389/ti.2022.10335},
  url={https://doi.org/10.3389/ti.2022.10335}
}

@book{Venugopal2023StatPearls,
  title={Biochemistry, C Peptide},
  author={Venugopal, Senthil K. and Mowery, Myles L. and Jialal, Ishwarlal},
  year={2023},
  publisher={StatPearls Publishing},
  address={Treasure Island (FL)},
  url={https://www.ncbi.nlm.nih.gov/books/NBK538129/}
}

@article{Lam2022BETA2,
  author  = {Lam, Anna and Oram, Richard A. and Forbes, Shareen and Olateju, Tolu and Malcolm, Andrew J. and Imes, Sharleen and Shapiro, A. M. James and Senior, Peter A.},
  title   = {Estimation of Early Graft Function Using the {BETA-2} Score Following Clinical Islet Transplantation},
  journal = {Transplant International},
  year    = {2022},
  volume  = {35},
  pages   = {10335},
  doi     = {10.3389/ti.2022.10335}
}

@article{Holt2021Type1Consensus,
  author    = {Holt, R. I. G. and DeVries, J. H. and Hess-Fischl, A. and Hirsch, I. B. and Kirkman, M. S. and Klupa, T. and Ludwig, B. and N{\o}rgaard, K. and Pettus, J. and Renard, E. and Skyler, J. S. and Snoek, F. J. and Weinstock, R. S. and Peters, A. L.},
  title     = {The Management of Type 1 Diabetes in Adults: A Consensus Report by the American Diabetes Association (ADA) and the European Association for the Study of Diabetes (EASD)},
  journal   = {Diabetes Care},
  year      = {2021},
  volume    = {44},
  number    = {11},
  pages     = {2589--2625},
  doi       = {10.2337/dci21-0043},
  url       = {https://doi.org/10.2337/dci21-0043},
  eprint    = {https://diabetesjournals.org/care/article-pdf/44/11/2589/648559/dci210043.pdf},
  note      = {PMID: 34593612}
}

\appendix
\section*{Appendix A: Simulation Settings}
In all simulation settings, we generated variables from the following distributions:
\begin{equation*}
    \begin{split}
        \mathbf{X}_{1,\cdots,5} \mathop{\sim}\limits^{iid} \mathcal{N}(0,1);\ \mathbf{X}_{6,\cdots,10} \mathop{\sim}\limits^{iid} \mathcal{U}(-1,1);\ \mathbf{X}_{11,\cdots,15} \mathop{\sim}\limits^{iid} \operatorname{Bernoulli}(0.5)
    \end{split}
    \label{eq:simsetting}
\end{equation*}

In the external data, outcome and variables have the following association:
\begin{equation*}
    \begin{split}
        Y^{E}\mid \boldsymbol{\beta}^{E},\sigma^2 \sim \mathcal{N}(0.5X_1 + 1.5X_2 - 0.6X_6 - 1.5X_7 + 0.4X_{11} + 1.2X_{12}, \sigma^2)
    \end{split}
    \label{eq:simsetting}
\end{equation*}

In the internal data, outcome and variables have the following association:
\begin{equation*}
    \begin{split}
        Y^{I}\mid &\ \boldsymbol{\beta}^{I},\sigma^2 \sim \mathcal{N}(0.5X_1 + 1.5X_2 - 0.6X_6 - 1.5X_7 + 0.4X_{11} + 1.2X_{12}, \sigma^2)\ \ \text{(Identical Signals)}\\
        Y^{I}\mid &\ \boldsymbol{\beta}^{I},\sigma^2 \sim \mathcal{N}(0.8X_1 + 1.8X_2 - 1.0X_6 - 1.7X_7 + 0.3X_{11} + 1.4X_{12}, \sigma^2)\ \ \text{(Same Pattern)}\\
        Y^{I}\mid &\ \boldsymbol{\beta}^{I},\sigma^2 \sim \mathcal{N}(1.0X_2 + 1.6X_3 - 0.8X_7 - 1.2X_8 + 0.8X_{12} + 1.5X_{13}, \sigma^2)\ \ \text{(Partial Overlap)}\\
        Y^{I}\mid &\ \boldsymbol{\beta}^{I},\sigma^2 \sim \mathcal{N}(1.0X_3 + 1.6X_4 - 0.8X_8 - 1.2X_9 + 0.8X_{13} + 1.5X_{14}, \sigma^2)\ \ \text{(No Overlap)}
    \end{split}
    \label{eq:simsetting}
\end{equation*}

\section*{Appendix B: Additional Simulation Results}
We summarized the mean RMSE of each method with internal data sample size $n^{I}\in\{10,20,30\}$ in Figure \ref{fig:figS1}. In general, mean RMSE increases with internal data sample size decreases, and the relative ranking of RMSE of APSP remains the same or slightly changes with internal data sample sizes.

\begin{figure}
    \centering
    \includegraphics[width=1\linewidth]{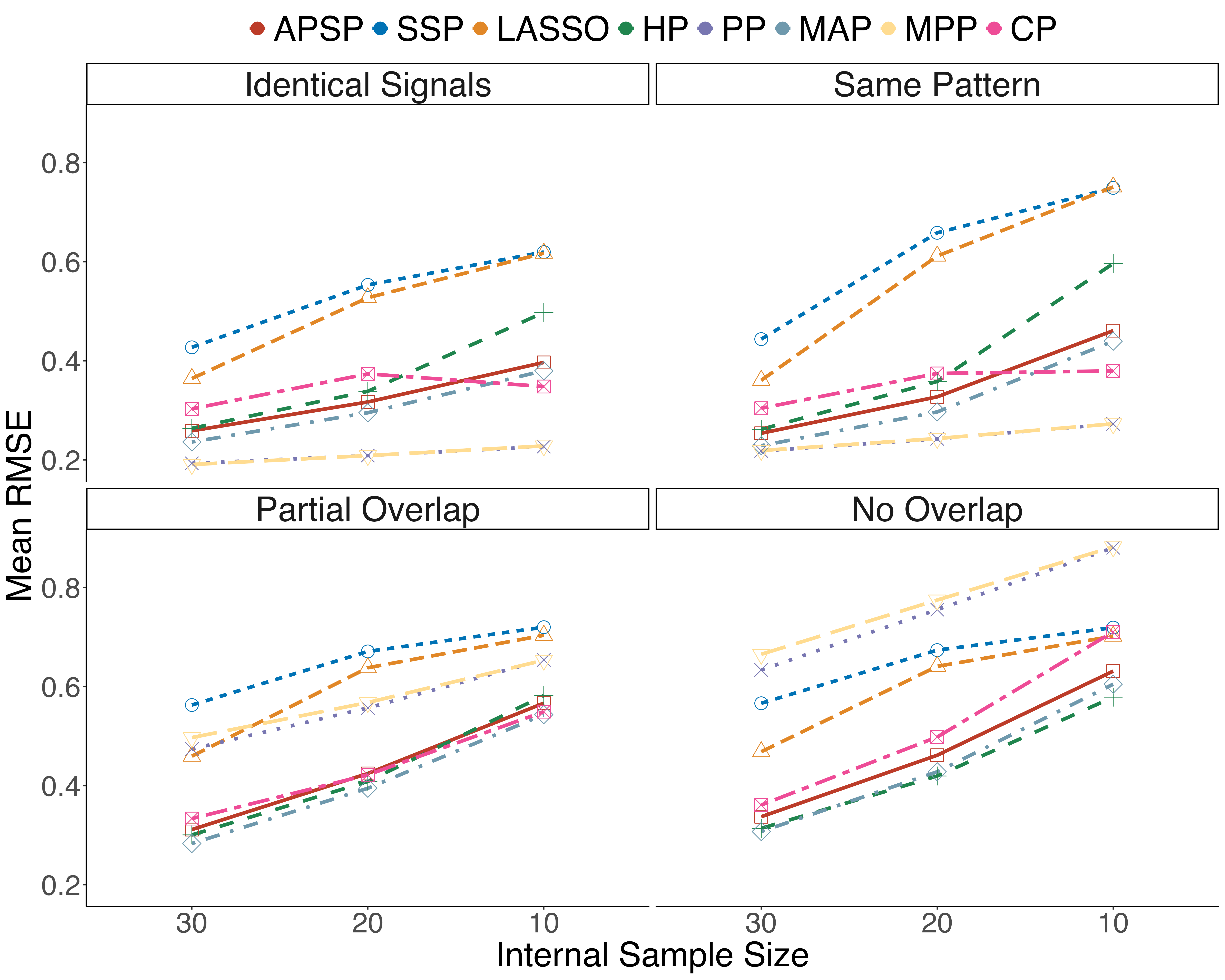}
    \caption{Root Mean Square Error (RMSE) of proposed method and competing methods under multiple scenarios. Different combinations of color, dot shape, and line type represent distiguish methods.
    }
    \label{fig:figS1}
\end{figure}

\begin{figure}
    \centering
    \includegraphics[width=1\linewidth]{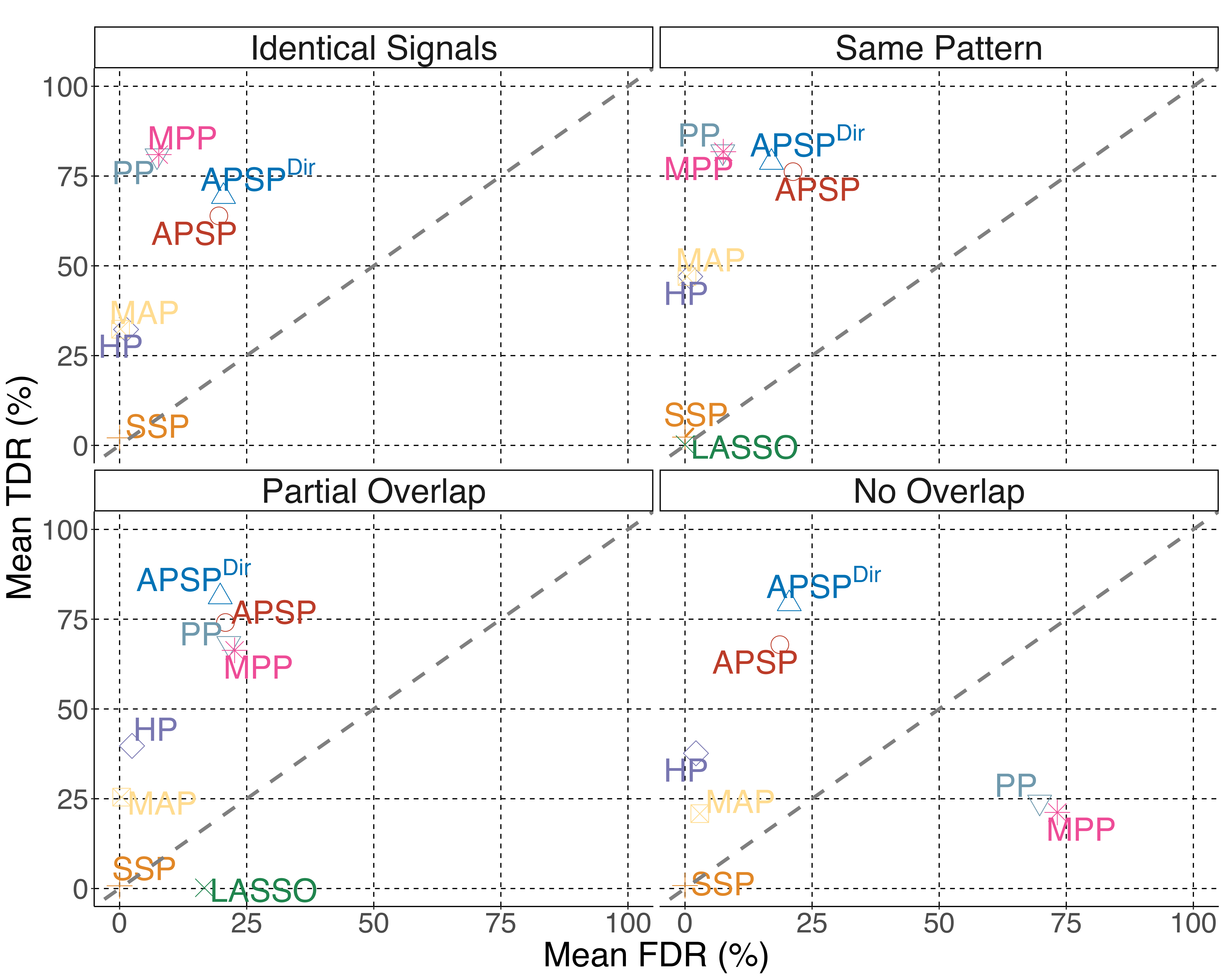}
    \caption{Scatter plot of TDR $(\%)$ and FDR $(\%)$ of $\text{APSP}^{\text{Dir}}$, APSP and competing methods under multiple scenarios with internal data sample size at 20. Smaller RMSE means better estimation of model parameters. High TDR and low FDR are favored, thus the methods locates at top-left is better.
    }
    \label{fig:figSB21}
\end{figure}

\begin{figure}
    \centering
    \includegraphics[width=1\linewidth]{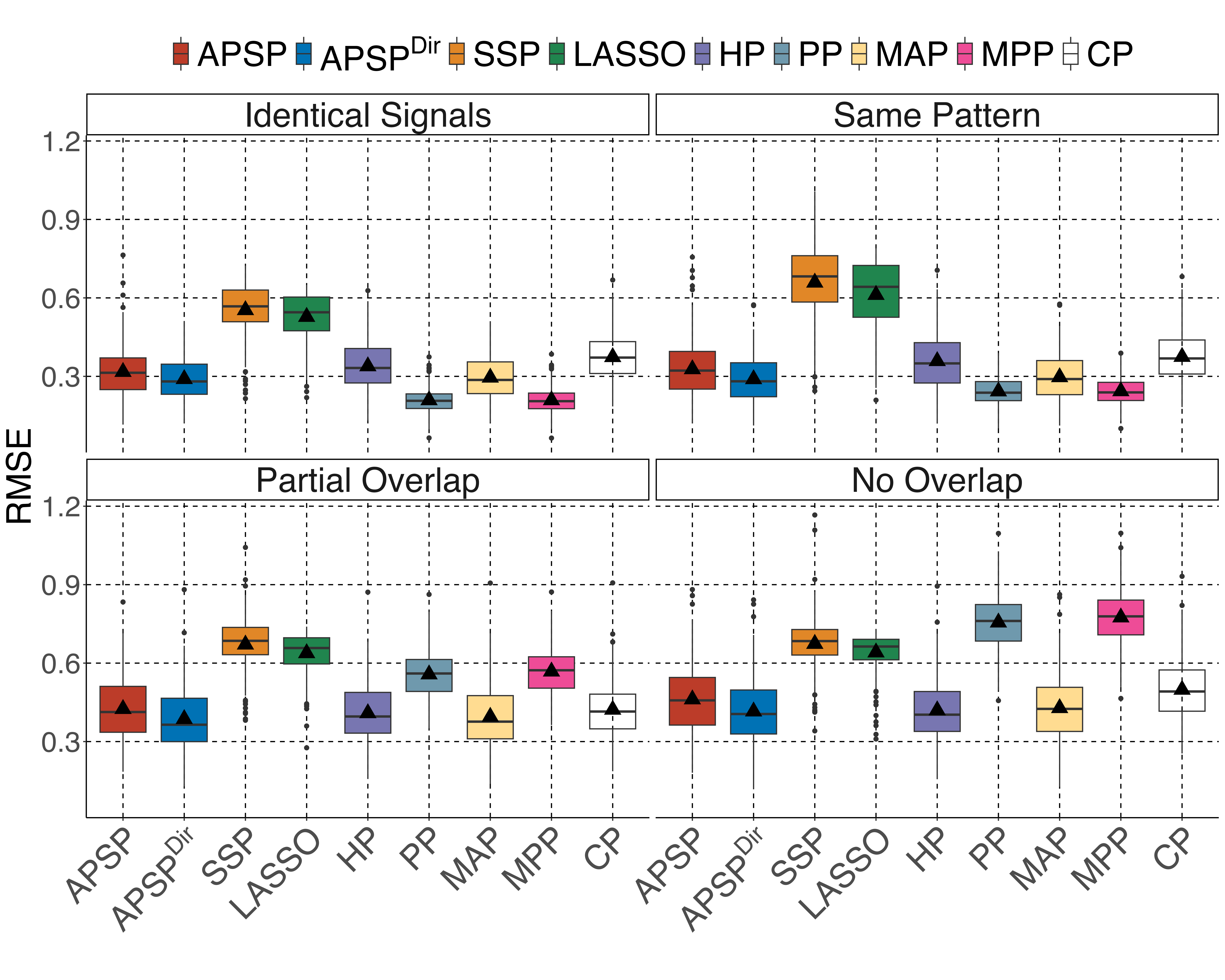}
    \caption{Root Mean Square Error (RMSE) of $\text{APSP}^{\text{Dir}}$, APSP and competing methods under multiple scenarios with internal data sample size at 20. Smaller RMSE means better estimation of model parameters.
    }
    \label{fig:fig2}
\end{figure}

\begin{figure}
    \centering
    \includegraphics[width=1\linewidth]{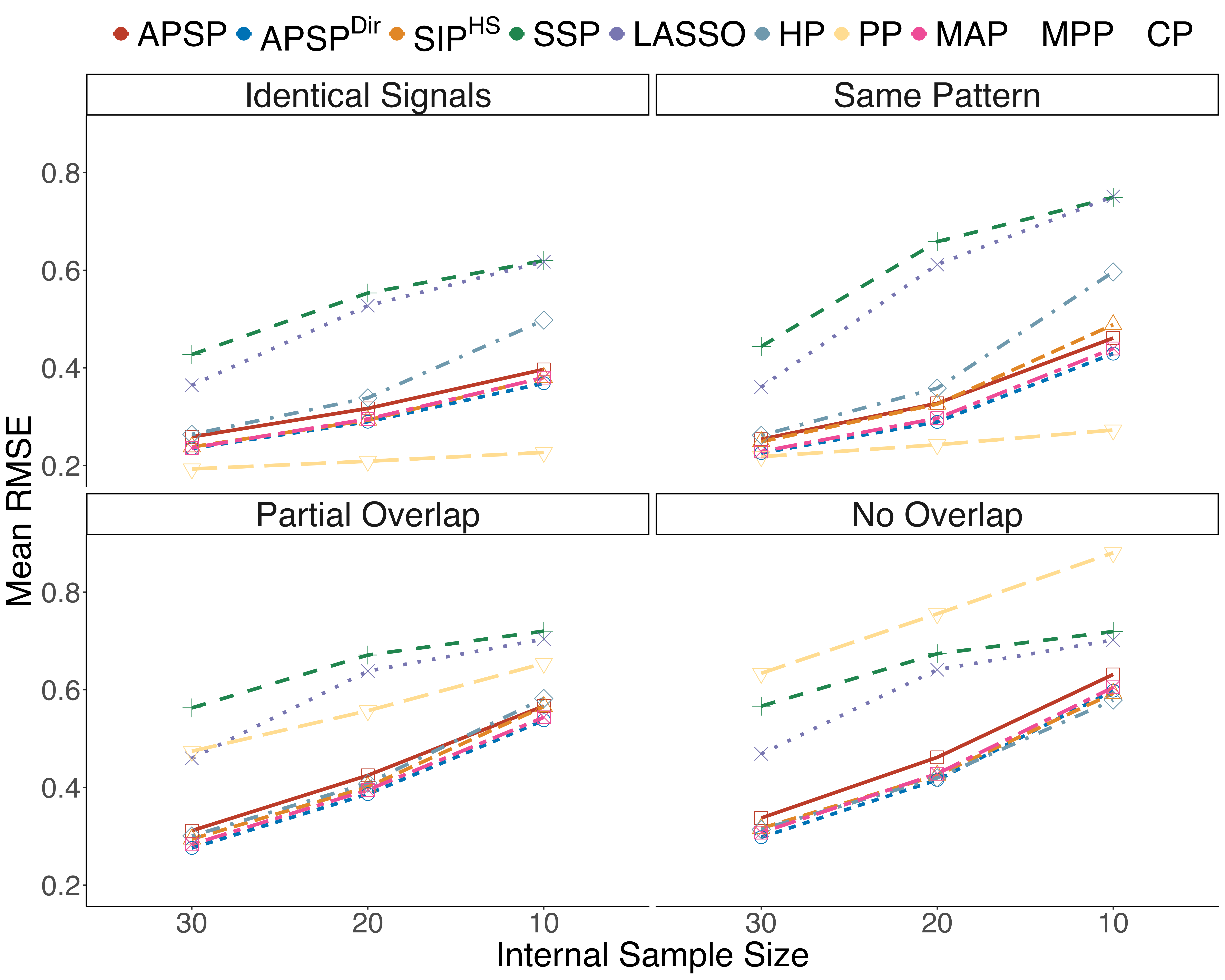}
    \caption{Root Mean Square Error (RMSE) of $\text{APSP}^{\text{Dir}}$, APSP and competing methods under multiple scenarios. Different combinations of color, dot shape, and line type represent distinguish methods.
    }
    \label{fig:figS1}
\end{figure}

\end{document}